\documentclass[conference]{IEEEtran}
\usepackage{cite}
\usepackage[dvips]{graphicx}
\usepackage[T1]{fontenc}
\usepackage[cmex10]{amsmath}
\usepackage{array}
\usepackage{theorem}
\usepackage[caption=false,font=footnotesize]{subfig}
\usepackage{fixltx2e}
\usepackage{url}
\usepackage{acronym}
\usepackage{comment}
\usepackage{soul,color}
\usepackage{multirow}
\usepackage{amsmath}
\usepackage{amssymb}
\usepackage{mathrsfs}
\usepackage{mathptmx}
\usepackage[scaled]{helvet}    
\usepackage{txfonts}

{\theorembodyfont{\rmfamily}
   }
{\theorembodyfont{\rmfamily}
   }
{\theorembodyfont{\rmfamily}
   }
{\theorembodyfont{\rmfamily}
   }

{\theorembodyfont{\rmfamily}
   }
{\theorembodyfont{\rmfamily}
   }

\acrodef{LDPC}{low-density parity-check}
\acrodef{BER}{bit error rate}
\acrodef{FER}{frame error rate}
\acrodef{CER}{codeword error rate}
\acrodef{BLER}{block error rate}
\acrodef{UB}{union bound}
\acrodef{QAM}{quadrature amplitude modulation}
\acrodef{ML}{maximum likelihood}
\acrodef{BI-DMS}{binary-input discrete memoryless symmetric}
\acrodef{BEC}{binary erasure channel}
\acrodef{CRC}{cyclic redundancy check}
\acrodef{WEF}{weight enumerating function}
\acrodef{AWEF}{average weight enumerating function}
\acrodef{IOWEF}{input output weight enumerating function}
\acrodef{MDS}{maximum distance separable}
\acrodef{SC}{successive cancellation}
\acrodef{BCH}{Bose-Chaudhuri-Hocquenghem}
\acrodef{BEP}{block error probability}
\acrodef{IRWEF}{input redundancy weight enumerating function}

\begin{document}

\title{On the Error Probability of Short Concatenated Polar and Cyclic Codes with Interleaving}
\author{\IEEEauthorblockN{Giacomo Ricciutelli, Marco Baldi, Franco Chiaraluce}
\IEEEauthorblockA{Universit\`a Politecnica delle Marche,\\
Ancona, Italy\\
Email: g.ricciutelli@pm.univpm.it, \{m.baldi, f.chiaraluce\}@univpm.it}
\and
\IEEEauthorblockN{Gianluigi Liva}
\IEEEauthorblockA{German Aerospace Center (DLR),\\
Wessling, Germany\\
Email: gianluigi.liva@dlr.de}}

\maketitle

\pagestyle{empty}
\thispagestyle{empty}

\begin{abstract}
In this paper, the analysis of the performance of the concatenation of a short polar code with an outer binary linear block code is addressed from a distance spectrum viewpoint. The analysis targets the case where an outer cyclic code is employed together with an inner systematic polar code. A concatenated code ensemble is introduced placing an interleaver at the input of the polar encoder. The introduced ensemble allows deriving bounds on the achievable error rates under maximum likelihood decoding, by applying the union bound to the (expurgated) average weight enumerators. The analysis suggests the need of careful optimization of the outer code, to attain low error floors.
\end{abstract}

\section{Introduction}
\label{sec:Intro}
Polar codes \cite{Arikan2009,stolte2002rekursive} provably achieves the capacity of \ac{BI-DMS} channels by using the (low complexity) \ac{SC} decoding algorithm, in the limit of infinite block length. At short block lengths polar codes under \ac{SC} decoding tend to exhibit a poor performance. In \cite{Tal2015} it was suggested that such a behavior might be due, on one hand, to an intrinsic weakness of polar codes and, on the other hand, to the sub-optimality of \ac{SC} decoding w.r.t. \ac{ML} decoding. An improved decoding algorithms were proposed in \cite{Tal2015, Arikan2009b, Trifonov2012}, while the structural properties of  polar codes (e.g., their distance properties) were studied, among others,  in \cite{Korada2010, Guo2013, Valipour2013, Liu2014, Li2014, Bardet2016}. 
The minimum distance properties of polar codes can be improved by resorting to concatenated schemes like the one of \cite{Tal2015}, where the concatenation of polar codes with an outer \ac{CRC} code is considered. 
This solution, together with the use of the list decoding algorithm of \cite{Tal2015}, allows short polar codes to become competitive against other families of codes \cite{Li2012, Niu2012, Bhatia2015,  Wu2016, Liva2016}.\footnote{The on-going 3GPP standardization group is considering the adoption of short polar codes with an outer \ac{CRC} for the uplink control channel of the upcoming 5th generation mobile standard \cite{Polar5G}.}

A theoretical characterization of the performance of concatenated \ac{CRC}-polar codes is still an open problem. Furthermore, for a fixed code length, a concatenated scheme can be realized with several combinations of the component codes parameters (e.g., one may choose various \ac{CRC} polynomials and polar codes designed for different target signal-to-noise ratios).

In this paper, we provide an analysis of the concatenation of polar codes with binary cyclic outer codes. The analysis is carried out by introducing concatenated ensembles and by deriving \ac{AWEF} by following the well-known uniform interleaver approach \cite{Benedetto1996}. In the analysis, the knowledge of the \ac{IOWEF} of the inner polar code is required. However, the polar code \ac{IOWEF} calculation through analytical methods is still an unsolved problem. Hence, we restrict our attention to short, high-rate polar codes for which the problem can be solved through a pragmatic approach. More precisely, we consider the dual code of the selected polar code and then we find its \ac{IOWEF} by listing the codewords.
Subsequently, by using the generalized MacWilliams identity \cite{MacWilliams1977}, we obtain the \ac{IOWEF} of the original polar code. The \ac{WEF} of the outer cyclic code, instead, is computed by following the method presented in \cite{Wolf1988}.

By adopting the uniform interleaver approach, we subsume the existence of an interleaver between the inner and the outer code, and obtain the average performance of an ensemble composed by the codes obtained by selecting all possible interleavers. 
Our analysis shows that the performance of the concatenated scheme with and without interleaver (as proposed in \cite{Tal2015}) may differ substantially. Similarly, by considering both \ac{CRC} and \ac{BCH} outer codes, we show that the choice of the outer code plays an important role in the short block length regime.

The paper is organized as follows. Notation and definitions are introduced in Section \ref{sec:Definitions}. The distance spectrum analysis of the concatenated scheme is discussed in Section \ref{sec:UpperUnionBound}. Numerical results are reported in Section \ref{sec:NumericalResults}. Finally, Section \ref{sec:Conclusion} concludes the paper.

\section{Notation and Definitions}
\label{sec:Definitions}

Let $w_H(\cdot)$ be the Hamming weight of a vector.
We denote as $N$ and $K$ the outer cyclic code length and dimension, respectively, while $n$ and $k$ identify the same parameters for the inner polar code. 
Therefore, the code rates of the outer and inner code are $R_O=\frac{K}{N}$ and $R_I=\frac{k}{n}$, respectively. 
The two codes can be serially concatenated on condition that $N=k$, thus the overall code rate of the concatenated code is $R=\frac{K}{n}$.
Given a binary linear code $\mathcal{C}(n,k)$, its \ac{WEF} is defined as \cite{MacWilliams1977}
\begin{equation}
A_\mathcal{C}(X)=\sum_{i=0}^NA_iX^i
\label{eq:WEF}
\end{equation}
where $A_i$ is the number of codewords $\mathbf{c}$ with $w_H(\mathbf{c})=i$. In this work we focus on systematic polar codes (the reason of this choice will be discussed later) so the first $k$ bits of a codeword $\mathbf{c}$ coincide with the information vector $\mathbf{u}$, yielding $\mathbf{c}=(\mathbf{u}|\mathbf{p})$ with $\mathbf{p}$ being the parity vector.
The \ac{IOWEF} of a code $\mathcal{C}(n,k)$ is
\begin{equation}
A_\mathcal{C}^\textnormal{IO}(X, Y)=\sum_{i=0}^k\sum_{\omega=0}^nA_{i,\omega}^\textnormal{IO}X^{i}Y^\omega
\label{eq:IOWEF}
\end{equation}
where $A_{i,\omega}^\textnormal{IO}$ is the multiplicity of codewords $\mathbf{c}$ with $w_H(\mathbf{u})=i$ and $w_H(\mathbf{c})=\omega$.
The enumeration of the codeword weights entails a large complexity even for small code dimensions.
In order to overcome this problem and obtain the \ac{IOWEF} of the considered polar codes, we focus on short, high-rate polar codes and we exploit the generalized MacWilliams identity \cite{MacWilliams1977}. 
This approach was followed also for cyclic codes, for example, in \cite{Wolf1988} to compute the \ac{WEF} of several \acp{CRC}.
Denote by $\mathcal{C}^\bot$ the dual code of $\mathcal{C}$. Given the dual code \ac{WEF} $A_{\mathcal{C}^\bot}(X)$, we can express the original code \ac{WEF} $A_{\mathcal{C}}(X)$ as \cite{MacWilliams1977}
\begin{equation}
A_{\mathcal{C}}(X)=\frac{\left(1+X\right)^n}{\left|\mathcal{C}^\bot\right|}A_{\mathcal{C}^\bot}\left(\frac{1-X}{1+X}\right)
\label{eq:MacwilliamsWEF}
\end{equation}
where $\left|\mathcal{C}^\bot\right|$ is the cardinality of the dual code.
When the \ac{IOWEF} is of interest, a significant reduction of the computational cost can be achieved by considering systematic codes. For such reason, in this work we have used only systematic inner polar codes. In the case of a systematic code $\mathcal{C}(n,k)$, it is convenient to derive the \ac{IOWEF} from the \ac{IRWEF} $A_\mathcal{C}^\textnormal{IR}(x, X, y, Y)$ defined as
\begin{equation}
A_\mathcal{C}^\textnormal{IR}(x, X, y, Y)=\sum_{i=0}^k\sum_{p=0}^rA_{i,p}^\textnormal{IR}x^{k-i}X^iy^{r-p}Y^p
\label{eq:IRWEF}
\end{equation}
where $A_{i,p}^\textnormal{IR}$ is the multiplicity of codewords $\mathbf{c}$ with $w_H(\mathbf{u})=i$ and $w_H(\mathbf{p})=p$, with $w_H(\mathbf{c})=i+p$ and $n=k+r$.
Hence, starting from the \ac{IRWEF} of the dual code $A_\mathcal{C^\bot}^{IR}(x,X, y,Y)$, we have \cite{Weiss2001,Chiaraluce2004}
\begin{align}
&A_{\mathcal{C}}^\textnormal{IR}(x, X, y, Y)= \nonumber \\
&\frac{1}{\left|\mathcal{C}^\bot\right|}A_{\mathcal{C}^\bot}^\textnormal{IR}(x+X,x-X, y+Y, y-Y).
\label{eq:MacwilliamsIRWEF}
\end{align}
Then, the \ac{IOWEF} is obtained as
\begin{equation}
A_\mathcal{C}^\textnormal{IO}(X, Y)=A_\mathcal{C}^\textnormal{IR}(1, XY, 1, Y).
\label{eq:IRWEFtoIOWEF}
\end{equation}

\section{Union Bound of the Average Block Error Probability of the Concatenated Scheme}
\label{sec:UpperUnionBound}
Given an ensemble of binary linear codes $\mathscr{C}(n,k)$, the expected \ac{BEP} $P_B$ of a random code $\mathcal{C} \in \mathscr{C}(n,k)$ under \ac{ML} decoding over a \ac{BEC} with erasure probability $\epsilon$ can be upper bounded as \cite{Liva2013}
\begin{align}
& \mathbb{E}\left[P_B\left(\mathcal{C},\epsilon\right)\right] \leq P_B^{(s)}(n,k,\epsilon) \nonumber \\
&+\sum_{e=1}^k \binom{n}{e} \epsilon^e(1-\epsilon)^{n-e} \min\left\{1,\sum_{\omega=1}^e\binom{e}{n}\frac{\bar{A}_\omega}{\binom{n}{\omega}}\right\}
\label{eq:UpperUnionBound}
\end{align}
where $\bar{A}_\omega=\mathbb{E}\left[{A}_\omega\left(\mathcal{C}\right)\right]$ is the average multiplicity of codewords with $w_H(\mathbf{c})=\omega$ and $P_B^{(s)}$ represents the \ac{BEP} of an ideal \ac{MDS} code, with parameters $n$ and $k$.
The \ac{UB} in \eqref{eq:UpperUnionBound} is applicable to every code ensemble whose expected \ac{WEF} is known. In order to use \eqref{eq:UpperUnionBound}, the \ac{AWEF} of the concatenated code ensemble must be known.

When dealing with concatenated codes, it is commonplace to consider a general setting including an interleaver between the inner and the outer codes. The concatenated code ensemble is hence given by the codes obtained by selecting all possible interleavers.
The special case without any interleaver can then be modeled as an identity interleaver.
From \cite{Benedetto1996}, the \ac{AWEF} of a concatenation formed by an inner polar code and an outer cyclic code can be obtained from the cyclic code \ac{WEF} and the polar code \ac{IOWEF} as
\begin{equation}
\bar{A}_\omega=\sum_{i=0}^N\frac{A_i^\textnormal{out} \cdot A_{i,\omega}^\textnormal{IO,in}}{\binom{N}{i}}
\label{eq:UniformInterleaver}
\end{equation}
where $A_i^\textnormal{out}$ is the weight enumerator of the outer code and $A_{i,\omega}^\textnormal{IO,in}$ is the input-output weight enumerator of the inner code (we remind that the average multiplicities resulting from \eqref{eq:UniformInterleaver} are, in general, real numbers).

The ensemble $\mathscr{C}(n,k)$ contains the codes generated by all possible interleavers. Thus, also bad codes (i.e., characterized by bad error rate performance) belong to the ensemble. It is clear that the bad codes adversely affect the \ac{AWEF} obtained through \eqref{eq:UniformInterleaver} causing a too pessimistic estimate of the error probability obtained through \eqref{eq:UpperUnionBound}, with respect to that achieved by properly designed codes. A simple way to overcome this issue is to divide $\mathscr{C}(n,k)$ into the bad and good code subsets, and then derive the \ac{AWEF} only of good codes through the expurgated $\mathscr{C}(n,k)$ \cite{Bennatan2004}. In fact, $\bar{A}_\omega= \xi \bar{A}_\omega^g + (1- \xi) \bar{A}_\omega^b$, where $\bar{A}_\omega^g$ and $\bar{A}_\omega^b$ denote the good and the bad codes ensemble, respectively. In this work we have assumed $\xi=0.99$, hence at least one code belongs to the good codes ensemble.
Therefore, when the expurgation method is adoptable (i.e., the first term of the \ac{AWEF} has a codewords multiplicity less than $\xi$), through \eqref{eq:UpperUnionBound} and \eqref{eq:UniformInterleaver} the average performance of the good codes subset is derived.

Studying the dual codes and exploiting MacWilliams identities allow considerable reductions in complexity of exhaustive analysis as long as the original code rate is sufficiently large.
This will be the case for the component codes considered next, which are characterized by $R_O, R_I > \frac{1}{2}$.
Therefore, in our case \eqref{eq:MacwilliamsWEF} and \eqref{eq:IRWEFtoIOWEF} can effectively be exploited to calculate $A_i^\textnormal{out}$ and $A_{i,\omega}^\textnormal{IO,in}$ in \eqref{eq:UniformInterleaver}. 

\section{Code Examples}
\label{sec:NumericalResults}

In this section we consider several examples of polar-cyclic concatenated codes and assess their performance through the approach described in the previous sections. 
Our focus is on short, high rate component codes, which allow to perform exhaustive analysis of their duals in a reasonable time.
Our results are obtained considering a \ac{BEC} with erasure probability $\epsilon$.
As known, in the polar code construction a fixed value of the error transition probability is considered. 
Therefore, as usual in literature, in each of the following examples the polar code is designed by using $\epsilon=0.3$. 
All performance curves provided next are obtained through \eqref{eq:UpperUnionBound}. 
We consider codes with $n=64$ bits and a \ac{CRC} or a \ac{BCH} outer code, with $N$ according to the polar code dimension.
In this work, we have used a \ac{CRC}$-8$ and a \ac{CRC}$-16$ with the following generator polynomials $g(x)$ \cite{Wolf1988}:
\begin{itemize}
\item \ac{CRC}$-8$: $g(x)=x^8+x^2+1$;
\item \ac{CRC}$-16-$CCITT: $g(x)=x^{16}+x^{12}+x^5+1$.
\end{itemize}
Instead, regarding the \ac{BCH} code, we study two \ac{BCH} codes that have the same redundancy as the \ac{CRC}-8 and \ac{CRC}-16 codes. Thus, for the same $R$ value, a performance comparison of the results achieved by using a \ac{CRC} or a \ac{BCH} outer code is feasible and fair. In order to keep the selected polar codes unchanged, we have considered shortened \ac{BCH} codes.

For the case without interleaver considered in \cite{Tal2015}, the generator matrix $G$ of the concatenated code can be obtained from the cyclic and polar code generator matrices $G_{C}$ and $G_{P}$, respectively, as 

\begin{equation}
G = G_{C} \cdot G_{P}.
\label{eq:G}
\end{equation}
When we instead consider the more general case with an interleaver between the inner and outer codes, we have 

\begin{equation}
 G = G_{C} \cdot I \cdot G_{P}
\label{eq:G_Interleaver}
\end{equation}
where $I$ is an $N \times N$ permutation matrix representing the interleaver. 
Considering all codes in $\mathscr{C}(n,k)$, \eqref{eq:UniformInterleaver} leads to their \ac{AWEF} and hence to the average performance in terms of \ac{UB}, according to the uniform interleaver approach.
In order to have an idea of the gap between this average performance and those of single codes in the ensemble, in each of the following figures we include the performance of codes randomly picked in $\mathscr{C}(n,k)$. 
Without changing the outer and inner code, this can be easily done by introducing a random interleaver between the cyclic code and the polar code in the concatenated scheme. Hence, in this case, the matrix $I$ in \eqref{eq:G_Interleaver} is a random permutation matrix.
For readability reasons, in addition to the solution without interleaver, in each of the following examples, we have considered only $25$ random interleavers; however, the obtained results allow to address some general conclusions. 
Moreover, when the results of the expurgated $\mathscr{C}(n,k)$ differs from that achieved by the \ac{AWEF}, its performance in terms of \ac{UB} is also considered.
The \ac{UB} of the considered polar code alone is also included as a reference. 
Clearly, the comparison between the performance of the polar code and those of the concatenated schemes is not completely fair at least because of the different code rates; however, in this way, the performance gain achieved through concatenation can be pointed out. 

\begin{figure}[!t]
\begin{centering}
\includegraphics[width=90mm,keepaspectratio]{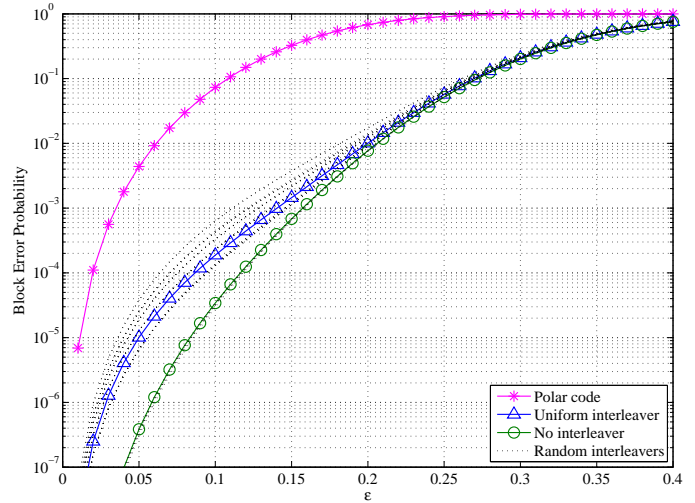}
\caption{Estimated performance of concatenated codes with and without interleaver composed by a (64,48) polar code and a \ac{CRC}$-8$ code over the \ac{BEC} under \ac{ML} decoding. Performance of the (64,48) polar code alone is also reported.}
\label{fig:Polar64-48-CRC8}
\end{centering}
\end{figure}

\begin{figure}[!t]
\begin{centering}
\includegraphics[width=90mm,keepaspectratio]{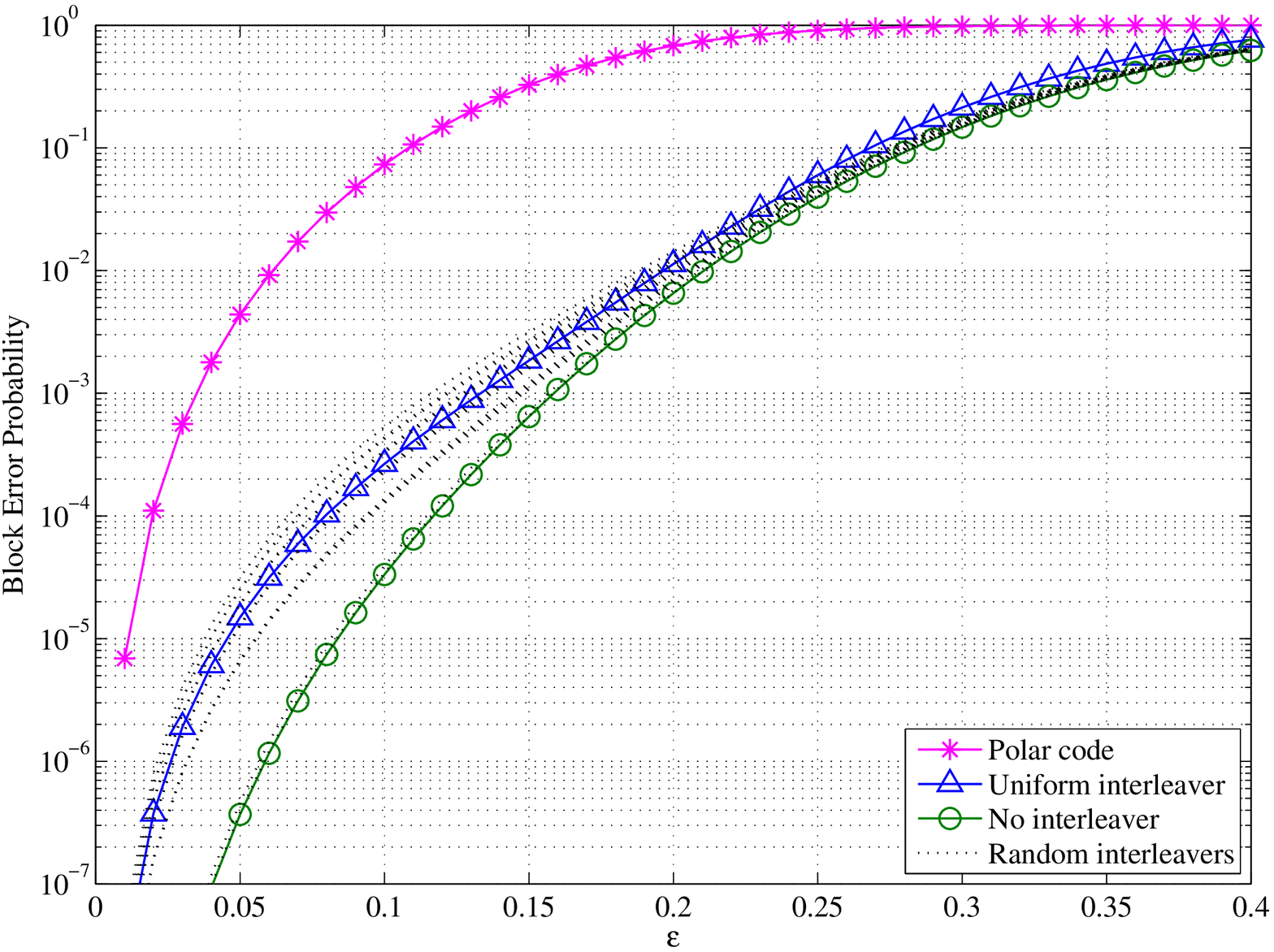}
\caption{Estimated performance of concatenated codes with and without interleaver composed by a (64,48) polar code and a (48,40) shortened \ac{BCH} code over the \ac{BEC} under \ac{ML} decoding. Performance of the (64,48) polar code alone is also reported.}
\label{fig:Polar64-48-BCH48-40}
\par\end{centering}
\end{figure}

Figures \ref{fig:Polar64-48-CRC8} and \ref{fig:Polar64-48-BCH48-40} show the \ac{UB} of the concatenated scheme, with and without random interleaver, formed by a polar code with $R_I=0.75$ and an outer cyclic code with $R_O=0.83$ (i.e., $R=0.625$) in terms of \ac{BEP}. In Fig. \ref{fig:Polar64-48-CRC8} and Fig. \ref{fig:Polar64-48-BCH48-40} a \ac{CRC}-8 and a (48, 40) \ac{BCH} outer code is considered, respectively, the latter obtained by shortening the (255, 247) \ac{BCH} code. In both the examples, the result obtained through the \ac{AWEF} corresponds to an average behavior, while some interleaver configurations achieve a smaller \ac{BEP}. This trend is due to the different minimum distance $d_{\text{min}}$ of the codes. The results in Figs. \ref{fig:Polar64-48-CRC8} and \ref{fig:Polar64-48-BCH48-40} are very similar, showing that, for this particular case, there is no substantial difference in performance arising from the different type of outer code. This conclusion is also supported by the results in Tab. \ref{tab:MinimumDistance}, where the number of concatenated schemes with random interleaver corresponding to a minimum distance value is reported. From these figures we can observe the performance gain introduced by the concatenated scheme with respect to the (64,48) polar code used alone, that has $d_{\text{min}}=4$. In both the examples the \ac{AWEF} and the concatenated code without interleaver have $d_{\text{min}}=4$ and $d_{\text{min}}=6$, respectively.

\begin{table}[!t] 
\renewcommand{\arraystretch}{1.1}
\caption{Number of concatenated codes with random interleaver for a given minimum distance value}
\label{tab:MinimumDistance}
\centering
\begin{tabular}{|c|c|c|c|}
\hline
Concatenated scheme&$d_{\text{min}}=4$&$d_{\text{min}}=6$&$d_{\text{min}}=8$\\
\hline
\hline
polar(64,48) + CRC-8& 21 & 4 & - \\ 
\hline
polar(64,48) + BCH(48,40)& 23 & 2 & - \\
\hline
polar(64,48) + CRC-16& 1 & 8 & 16 \\
\hline
polar(64,48) + BCH(48,32)& - & - & 25\\
\hline
polar(64,56) + CRC-16& 16 & 9 & - \\
\hline
\end{tabular}
\end{table} 

\begin{figure}[!t]
\begin{centering}
\includegraphics[width=90mm,keepaspectratio]{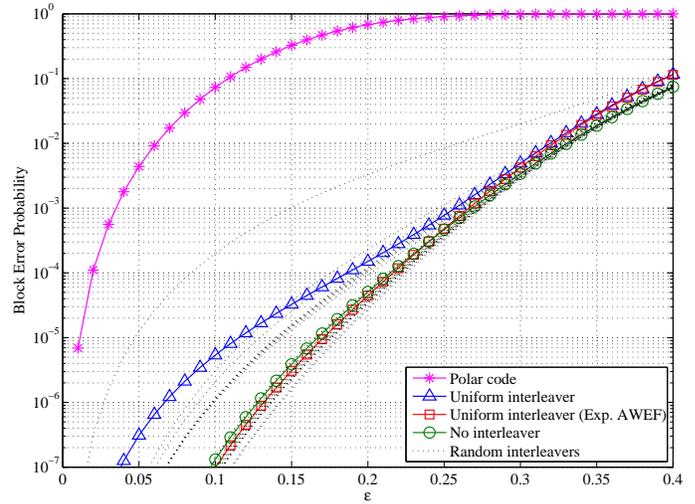}
\caption{Estimated performance of concatenated codes with and without interleaver composed by a (64,48) polar code and a \ac{CRC}$-16$ code over the \ac{BEC} under \ac{ML} decoding. Performance of the (64,48) polar code alone is also reported.}
\label{fig:Polar64-48-CRC16}
\par\end{centering}
\end{figure}

\begin{figure}[!t]
\begin{centering}
\includegraphics[width=90mm,keepaspectratio]{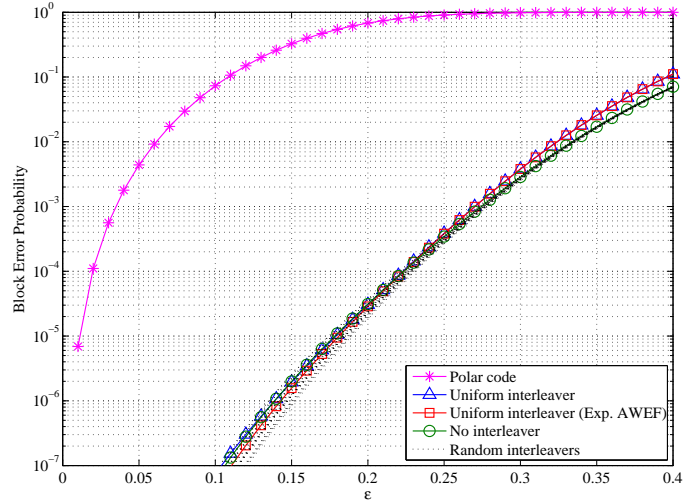}
\caption{Estimated performance of concatenated codes with and without interleaver composed by a (64,48) polar code and a (48,32) shortened \ac{BCH} code over the \ac{BEC} under \ac{ML} decoding. Performance of the (64,48) polar code alone is also reported.}
\label{fig:Polar64-48-BCH48-32}
\par\end{centering}
\end{figure}

In Fig. \ref{fig:Polar64-48-CRC16} and Fig. \ref{fig:Polar64-48-BCH48-32} the \ac{UB} of the concatenated codes, with and without interleaver, composed by a polar code with $R_I=0.75$ and an outer code with $R_O=0.66$ (i.e., $R=0.5$) in terms of \ac{BEP} is plotted. In Figs. \ref{fig:Polar64-48-CRC16} and \ref{fig:Polar64-48-BCH48-32} a \ac{CRC}-16 and a (48, 32) \ac{BCH} outer code is considered, respectively,  the latter obtained by shortening the (255, 239) \ac{BCH} code. Differently from the previous figures, the \ac{UB} obtained with the expurgated \ac{AWEF} is now available. As in Figs. \ref{fig:Polar64-48-CRC8} and \ref{fig:Polar64-48-BCH48-40}, the curve obtained through the \ac{AWEF} well describes the ensemble average performance, while we see that the curve corresponding to the expurgated \ac{AWEF} belongs to the group of best codes. Also in these cases the performance gap between the (64,48) polar code alone and the concatenated codes is remarkable. In both the examples the \ac{AWEF} has $d_{\text{min}}=6$, while the expurgated \ac{AWEF} and the concatenated scheme without interleaver have $d_{\text{min}}=8$. However, from Fig. \ref{fig:Polar64-48-BCH48-32}, we can observe that, on the contrary to Fig. \ref{fig:Polar64-48-CRC16}, all curves of concatenated codes are very close. In fact, for the case in Fig. \ref{fig:Polar64-48-BCH48-32} only the \ac{AWEF} results in $d_{\text{min}}=6$ but with a codewords multiplicity equal to 0.0336, instead the realizations of concatenated codes have $d_{\text{min}}=8$, as shown in Tab. \ref{tab:MinimumDistance}. 
Therefore, differently from the \ac{CRC}, in this case the use of a \ac{BCH} code is able to increase the minimum distance of the concatenated code (also for the solution without interleaver); thus, for this specific case, the \ac{BCH} code should be preferred to the \ac{CRC} code.

\begin{figure}[!t]
\begin{centering}
\includegraphics[width=90mm,keepaspectratio]{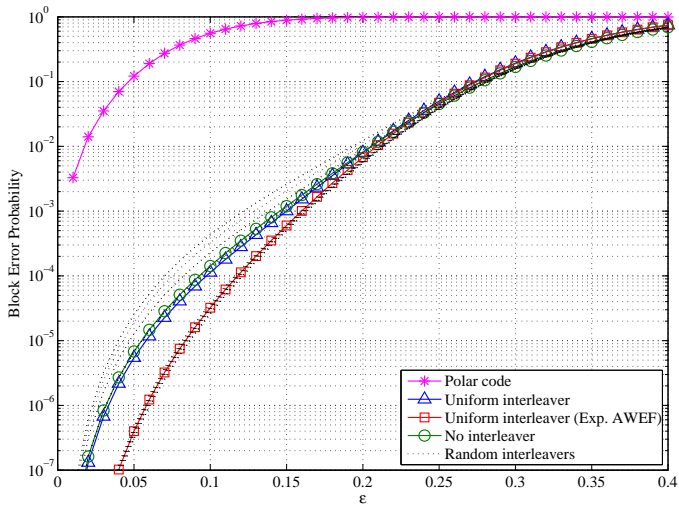}
\caption{Estimated performance of concatenated codes with and without interleaver composed by a (64,56) polar code and a \ac{CRC}$-16$ code over the \ac{BEC} under \ac{ML} decoding. Performance of the (64,56) polar code alone is also reported.}
\label{fig:Polar64-56-CRC16}
\par\end{centering}
\end{figure}

In all the previous figures the curve of the concatenated scheme without interleaver proposed in \cite{Tal2015} falls within the group of best performing codes.
This may lead to the conclusion that this configuration always produces a good result; in reality this trend is not preserved for any choice of the code parameters.
An example of the latter kind is shown in Fig. \ref{fig:Polar64-56-CRC16}, where a polar code with $R_I=0.875$ and a \ac{CRC}-16 code (i.e., $R_O=0.71$ and $R=0.625$) are considered. 
From the figure we observe that, in this case, the introduction of a random interleaver can improve the scheme without interleaving. Also for this example, Tab. \ref{tab:MinimumDistance} summarizes the number of concatenated schemes with interleaver for each value of the code minimum distance. In this case the polar code has $d_{\text{min}}=2$, while both the \ac{AWEF} and the concatenated scheme without interleaver have $d_{\text{min}}=4$. Instead, the expurgated \ac{AWEF} belongs to the group of good codes with $d_{\text{min}}=6$. We have found a similar result also by using the \ac{CRC}-8  code in place of the \ac{CRC}-16 but it is omitted for the sake of brevity. So, these counter examples (others can be found) clearly demonstrate that the use of a selected interleaver may be beneficial from the error rate viewpoint.

\section{Conclusion}
\label{sec:Conclusion}

In this paper, the analysis of the performance of the concatenation of a short polar code with an outer binary linear block code is addressed from a distance spectrum viewpoint.
The analysis is carried out for the case where an outer cyclic code is employed together with an inner systematic polar code. By introducing an interleaver  at the input of the polar encoder, we show that remarkable differences on the block error probability at low erasure probabilities can be observed for various permutations. The variations are due to the change in the overall concatenated code minimum distance (and minimum distance multiplicity) induced by the choice of the interleaver.
Bounds on the achievable error rates under maximum likelihood decoding are obtained by applying the union bound to the (expurgated) average weight enumerators.
The results point to the need of careful optimization of the outer code, at least in the short block length regime, to attain low error floors.


\end{document}